\documentclass[a4paper,11pt]{article}
\usepackage{jinstpub} 
\usepackage{lineno}
\usepackage{siunitx}

\title{\boldmath Measurement of ionization yield of low energy ions in low pressure CF$_{4}$ gas for dark matter searches}

\author[1]{Satoshi~Higashino}
\author[1]{, Wakako~Toyama}
\author[2]{, Takuya~Shiraishi}
\author[2]{, Yasushi~Hoshino}
\author[3]{, Tatsuhiro~Naka}
\author[1]{, Kentaro~Miuchi}

\affiliation[1]{Department of Physics, Kobe University, Kobe, Hyogo 657-8501, Japan}
\affiliation[2]{Department of Physics, Kanagawa University, Yokohama, Kanagawa, 221-8686, Japan}
\affiliation[3]{Department of Physics, Toho University, Narashino, Chiba, 274-8510, Japan}

\emailAdd{higashino@phys.sci.kobe-u.ac.jp}

\abstract{
Direction-sensitive direct dark matter search experiments have been conducted using gaseous detectors.
In spite of the long history of the study on the energy deposition of charged particles in materials, a full agreement between the measured results and theoretical predictions, especially in a low energy scale, are yet to be achieved.
It is thus important to measure the ionization yields of recoil nuclei for the experiments with gaseous detectors using an ionization charge readout scheme.
This study measured the ionization yield using a low-energy ion beam facility at Kanagawa University.
The ionization yields for fluorine ions with an energy range of 5~$\sim$~50~keV were measured using a dedicated proportional wire chamber filled with CF$_{4}$ gas at 0.06~atm.
The low‐energy ion injection scheme into a gaseous detector was established and the ionization yield for fluorine ions was obtained to be 0.45 at 30~keV with a moderate dependence on the ion energy.
}


\arxivnumber{2602.17126} 

\begin{document}
\maketitle
\flushbottom

\section{Introduction}
\label{sec:intro}

Direct searches of dark matter in a form of Weakly Interacting Massive Particles (WIMPs) with its directionality have been conducted in the world.
The major strategy of the directional search is to use the track reconstruction of a nuclear recoil caused by WIMPs.
Since the energy scale of nuclear recoils is at most tens of keV, their trajectories are very short.
Experiments such as NEWAGE~\cite{Shimada:2023vky}, DRIFT~\cite{DRIFT:2016utn} and DM-TPC~\cite{Ahlen:2010ub} have conducted direct dark matter searches using low pressure gaseous Time Projection Chambers (TPCs) by measuring the trajectory of nuclear recoils.
Particularly, gases containing fluorine nucleus (e.g. CF$_{4}$, CHF$_{3}$, and SF$_{6}$) are widely used for spin-dependent dark matter searches with directionality lead by NEWAGE, DM-TPC, MIMAC~\cite{Beaufort:2023vlb}, and CYGNO~\cite{Amaro:2022gub} experiments.

Along with the improvements of short-track detection technologies, studies of the detector response for low energy nuclear recoils with $\sim$~keV scale are also essential.
In spite of the long history of the study on the energy deposition of charged particles in materials, a full agreement between the measured results and theoretical predictions, especially in the low energy scale, are yet to be achieved.
One of the difficulties is that there are various methods to measure the amount of energy deposition, depending on detectors.
Several direction-sensitive dark matter experiments adopt ionization charge readout.
Low-energy nuclear recoils in the gas show high stopping power and are influenced by the nuclear stopping  in addition to the ionization. Thus the energy deposition is a combination of complex processes including the production of excitons, multiple vibrational modes, ion–electron recombination, and thermalization.
Since the yield of ionization depends on incident particles, materials, and their conditions, parameters are complicated for the calculation models such as the Lindhard theory~\cite{osti_4153115} and the SRIM~\cite{ZIEGLER20101818} calculation, which are widely used.
In case of dark matter experiments, usually energy calibrations are 
carried out using mono-energy radioisotope sources (gamma or alpha rays) or X-ray generators although the ionization yields in these processes are not the same as those for the nuclear recoils due to the difference of particle type. 
Therefore, it is important to measure the ionization yield of the nuclear recoils in each case.

Injecting ion beams directly into the gas is a powerful method for evaluating the ionization yield through direct measurements.
For example, the COMIMAC facility~\cite{Muraz:2016upt} established a low energy ion and electron beam injection scheme into a gas detector with an interface of a 1~$\mu$m hole between the beamline and the gaseous chamber.
In this work, we performed an ionization yield measurement injecting ion beams of known energies into a gaseous detector using an ion implantation system with an acceleration voltage of up to 200~kV at Kanagawa University in Japan.
We updated the ion implantation interface to adopt ion beam injection to a gas detector and established a low-energy ion beam injection scheme at Kanagawa University in Japan.
An ionization yield measurement for fluorine ions in low pressure CF$_{4}$ gas was performed with a dedicated proportional wire chamber.

Section~\ref{sec:setup} describes the setup of the ion beamline and the gaseous detector.
The result of the measurement is denoted in Section~\ref{sec:experiment} and Section~\ref{sec:conclusion} concludes this study.

\section{Setup}
\label{sec:setup}

The experimental setup for the measurement of the ionization yield for the ion beams with various energies is described in this section. 
Section~\ref{sec:ion_beam} describes the accelerator at Kanagawa University.
The mechanism separating the high-vacuum beamline from the gas volume of the detector is also described in this section.
Section~\ref{sec:detector} introduces the gaseous detector for this experiment.

\subsection{Low energy ion beamline}
\label{sec:ion_beam}

\begin{figure}[ht]
    \centering
    \includegraphics[width=1.0\textwidth]{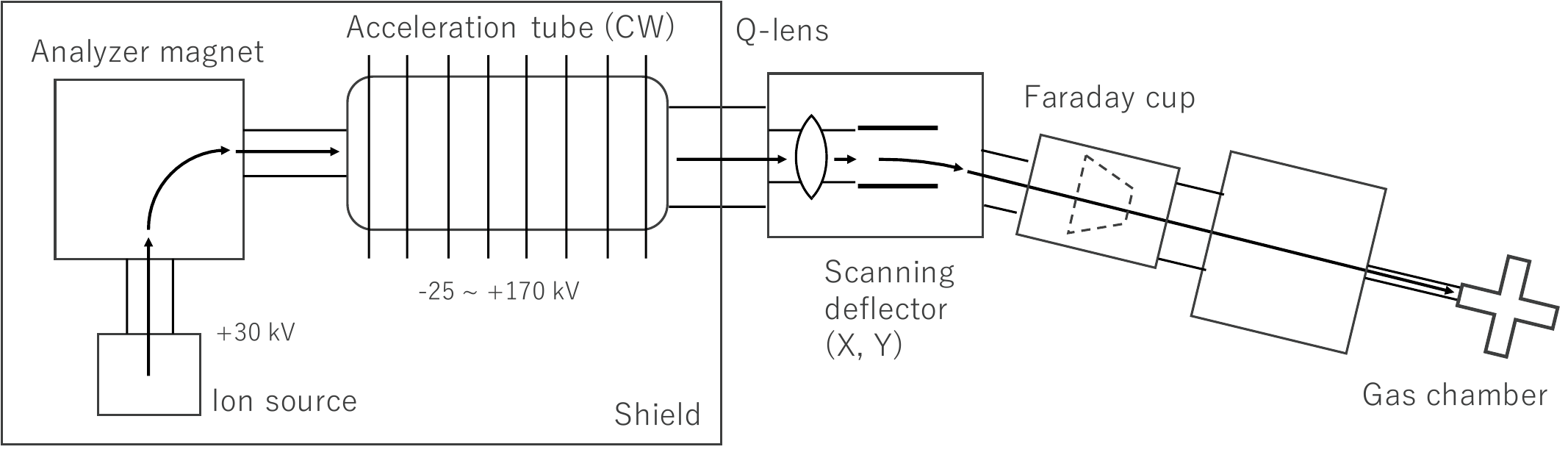}
    \caption{Schematic of the accelerator located at Kanagawa University.}
    \label{fig:accelerator}
\end{figure}

Figure~\ref{fig:accelerator} shows a schematic overview of the low-energy accelerator located at Kanagawa University. 
A Freeman-type ion source is employed as the ion source of the accelerator, in which the source gas is arc-discharged in an arc-chamber to stably form plasma using a hot filament and a magnetic field.
The generated ion species depend on the type of source gas; there have been successes in generating light ions to heavy ions (e.g. H$^{+}$, He$^{+}$, B$^{+}$, C$^{+}$, N$^{+}$, O$^{+}$, F$^{+}$, ... Xe$^{+}$).
Since fluorine nucleon recoil is assumed in various direction-sensitive dark matter search experiments, we selected F$^{+}$ ion beam in this study.
The F$^{+}$ ions are produced by decomposition of a SiF$_{4}$ source gas in the arc chamber.
The generated ions are extracted with a typical voltage of +30~kV and then an analyzing magnet selects the desired F$^{+}$ ions.
The selected ions are injected in an acceleration tube with applied voltages ranging from -25~kV to +170~kV with a Cockcroft-Walton (CW) generator. 
The beam energy is determined by the acceleration voltage and can therefore be tuned from 5~keV to 200~keV for monovalent ions.
The accelerated beam is shaped by a three-stage electrostatic quadrupole (Q-) lens system installed on the beamline.
Electrostatic scanning deflectors located downstream of the Q-lenses generate a uniform beam density by applying periodic fields at 833~Hz and 156~Hz in the $x$ and $y$ directions, respectively. 
The beamline is bent horizontally by $\ang{7}$ to prevent neutral particles from entering the detector.
The beam current is measured by a Faraday cup placed in the downstream of the scanning deflectors.
A gas detector is placed in the further downstream of the beamline connected via a thin stainless steel pipe with a 1/4-inch diameter.
The accelerated particles eventually pass through this thin pipe and enter the detector.

Although the beamline must be evacuated to the order of $10^{-4}$~Pa, the pressure inside the gas detector are typically maintained at $10^{5}$~Pa.
In order to separate these volumes with a pressure difference of almost the atmospheric pressure while ensuring the beam transport, a 10~$\mu$m-thick stainless steel film with a small injection hole was installed at the center of the exit port from the beam pipe.
The hole has a tapered structure with a diameter of 2.9~$\mu$m and 1.2~$\mu$m for the upstream and downstream of the beam direction, respectively.
Since the edge of the tapered beam hole could have affected the initial beam energy, we conducted a simulation study with Geant4 toolkit~\cite{AGOSTINELLI2003250}.
The simulation study ensured that events through the edge of tapered hole are negligible for this study.

\subsection{Detector}
\label{sec:detector}

\begin{figure}[ht]
    \centering
    \includegraphics[width=1.0\textwidth]{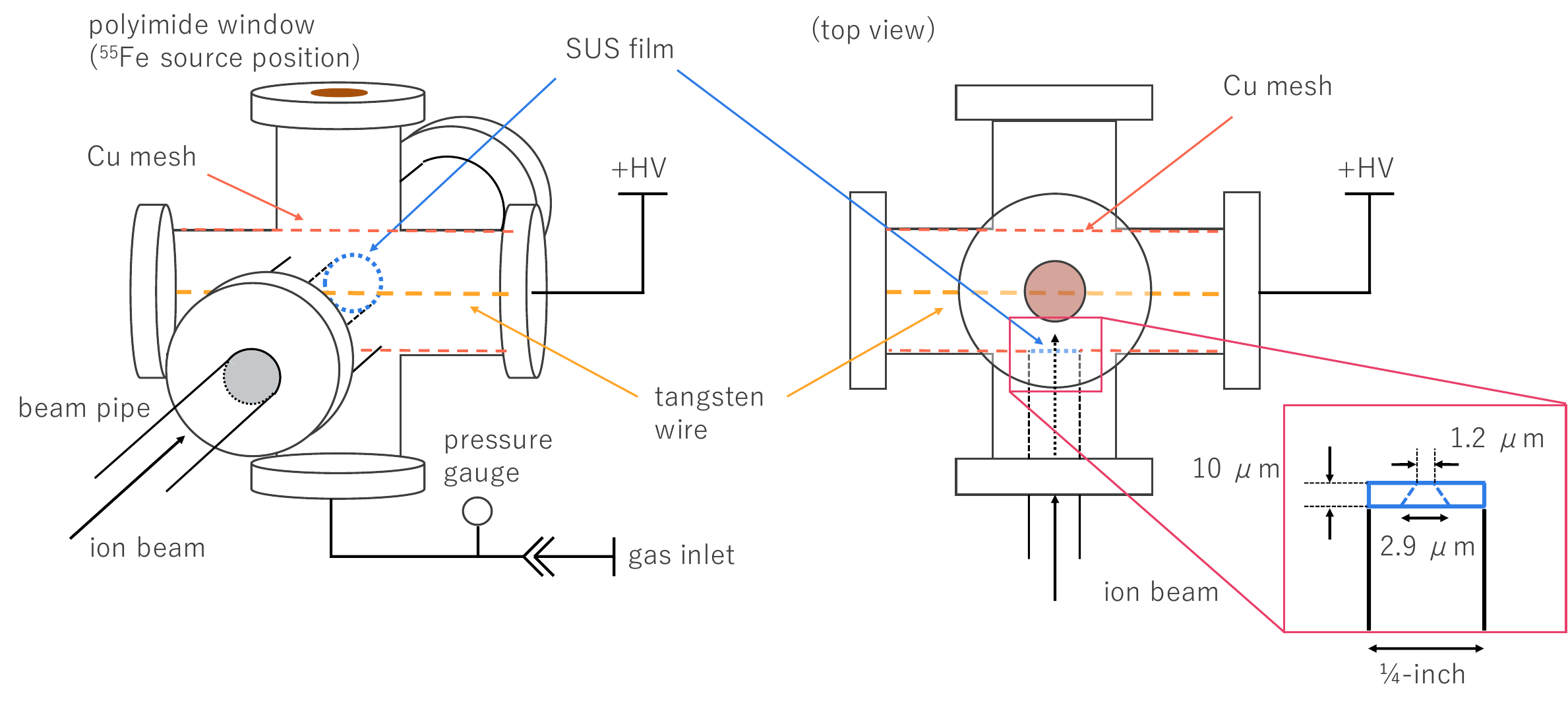}
    \caption{Schematic of the proportional wire chamber detector (left). Right schematic shows the top view of the detector and the tapered structure of the injection hole in the zoomed image.}
    \label{fig:detector}
\end{figure}

\begin{figure}[ht]
    \centering
    \includegraphics[width=0.8\textwidth]{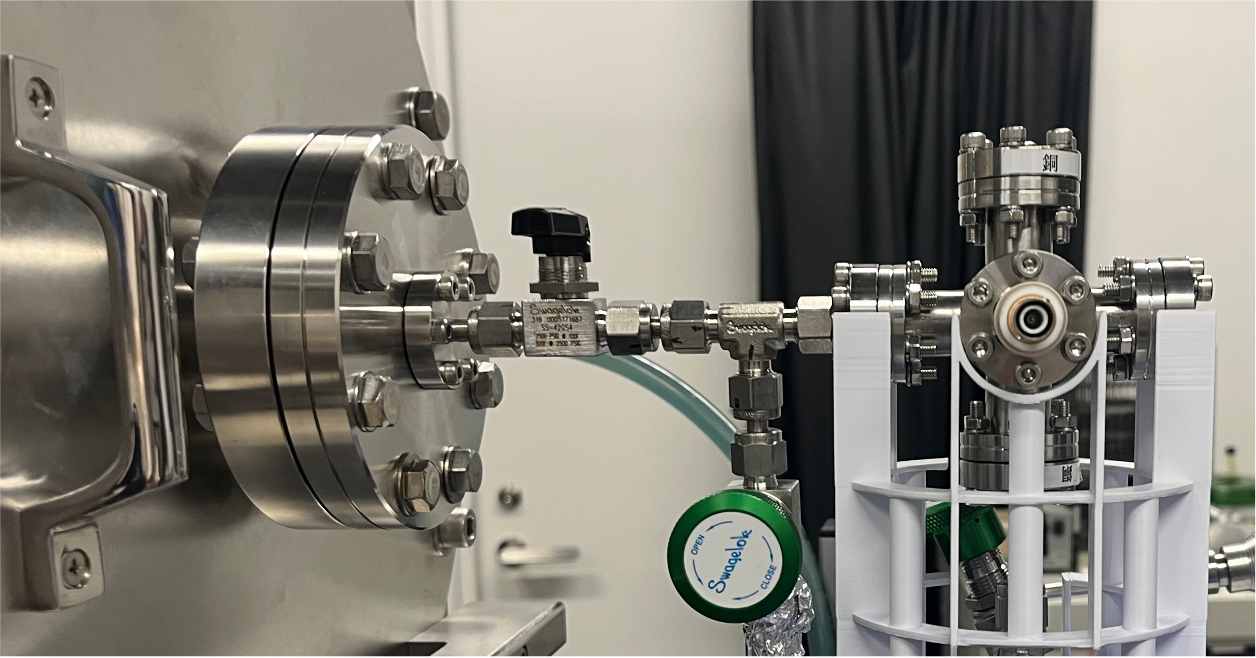}
    \caption{Picture of the proportional wire chamber detector (right-side) connected to the beamline (left-side).}
    \label{fig:det_pic}
\end{figure}

The energy deposition with ionization was measured using a proportional wire chamber.
Figure~\ref{fig:detector} shows the schematic of the wire chamber and Figure~\ref{fig:det_pic} shows the picture of the detector, including the interface to the beamline.
An ICF34 six-direction cross stainless steel pipe was customized to the wire chamber.
A single tungsten wire with a diameter of 30~$\mu$m was implemented in the center of the cross pipe as an anode electrode.
The inner diameter of the pipe was 15.8~mm and the chamber was connected to the ground as the cathode electrode.
To make a stable electrical field along with the anode wire, especially around the crossing part of the six direction cross pipe, a copper mesh cylinder was installed along with the wire.
The beam injection hole described in Section~\ref{sec:ion_beam} was connected on the surface of the copper mesh cylinder.
To avoid interference with the beam injection, a hole was made in the copper mesh cylinder around the beam entrance.
A flange with a 125~$\mu$m thick polyimide window was set on one side of the cross pipe to irradiate X-rays for the energy calibration.
The flange on the other side of the windows was connected to the gas inlet with a pressure monitor gage.
The chamber was filled with CF$_{4}$ at a pressure of 0.06~atm.

The anode wire was connected to a signal amplifier.
The charge was converted into the voltage with a charge-sensitive amplifier (Cremat Inc., CR-110) and then shaped with a dedicated shaper.
Waveforms of the signals were recorded using an ADALM2000 digitizer (Analog devices Inc.), which is capable of 100~MHz sampling with 12bit ADC ($\pm$2.5~V).
The charge of a signal was determined by the peak height of the ADC value where a pedestal value was subtracted.
The energy calibration scheme is described in Section~\ref{sec:calibration}.

\section{Experiment}
\label{sec:experiment}

A beam irradiation experiment was performed in the low-energy ion beamline at  Kanagawa University on 5th Sep. 2025.
The beam energy was scanned in 5~keV steps from 5~keV to 30~keV and in 10~keV steps from 30~keV to 50~keV, respectively.
The irradiation time for each energy was about 10~minutes in average and the total beam time was four hours including energy calibrations.
The beam current was monitored with the Faraday cup and confirmed to be stably operated at 150~nA.
The pressure in the beam pipe was as low as 4~$\times$~10$^{-5}$~Pa while the pressure in the gas chamber was kept at 0.06~atm within $\pm$~5\% uncertainty.
During the experiment, the anode current and the event rate were also monitored and were stable.
Section~\ref{sec:calibration} describes the method of the energy calibration using electron beams and an X-ray source.
Results of the beam experiment are shown in Section~\ref{sec:ion_yield}.

\subsection{Energy calibration}
\label{sec:calibration}

\begin{figure}[ht]
    \centering
    \includegraphics[width=0.7\textwidth]{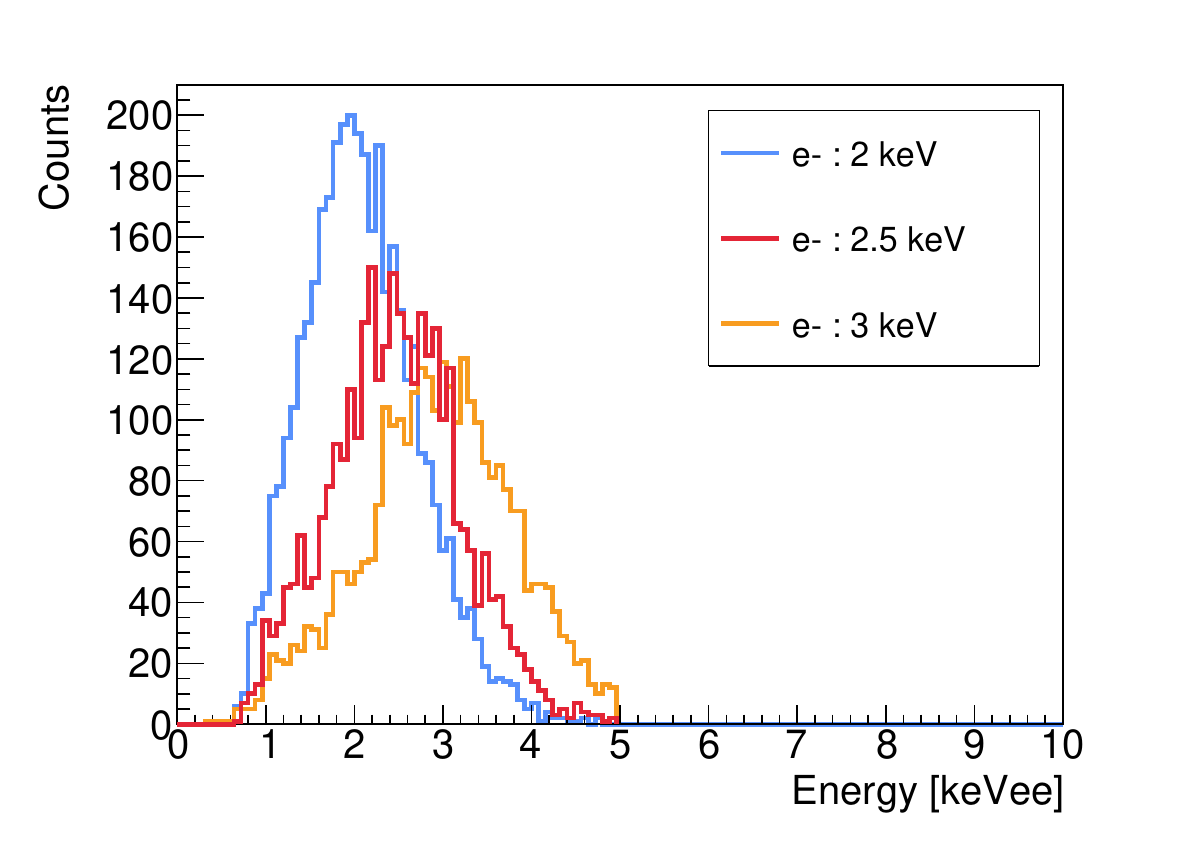}
    \caption{Energy spectra with an electron-gun for various electron energies.}
    \label{fig:electron_gun}
\end{figure}

Energy calibration for the ionization yield was carried out with two types of mono-energy sources.
An electron-gun (LK technologies, Model EG3000) was used as the primary source for the energy calibration.
The electron energy was tuned by the acceleration energy up to 3~keV.
In this calibration, the electron energies were set at 2.0, 2.5 and 3.0~keV.
Figure~\ref{fig:electron_gun} shows the energy spectra of the electron-gun measurements for these electron energies.
Each spectrum was fitted with a gaussian and the mean value of the gaussian was taken as the charge corresponding to the energy of the electron beam.
The unit of the energy was expressed as "electron-equivalent (ee)" energy (e.g. keV$_{\rm{ee}}$) in this paper.

\begin{figure}[ht]
    \centering
    \includegraphics[width=0.7\textwidth]{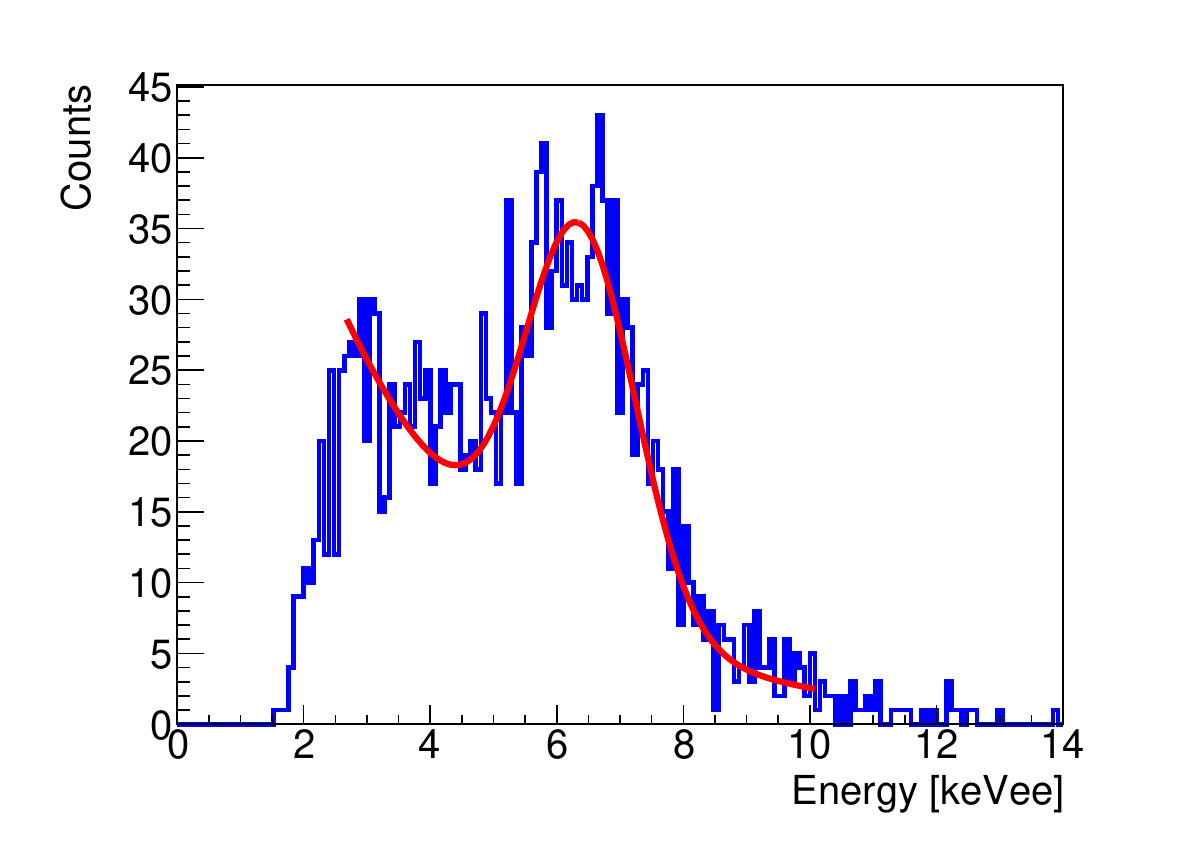}
    \caption{Energy spectrum with an $^{55}$Fe source. Measured spectrum (blue histogram) was fitted with a sum function of exponential and gaussian (red line). The energy scale was calibrated by the electron-gun measurements.}
    \label{fig:55Fe_calib}
\end{figure}

\begin{figure}[ht]
    \centering
    \includegraphics[width=0.7\textwidth]{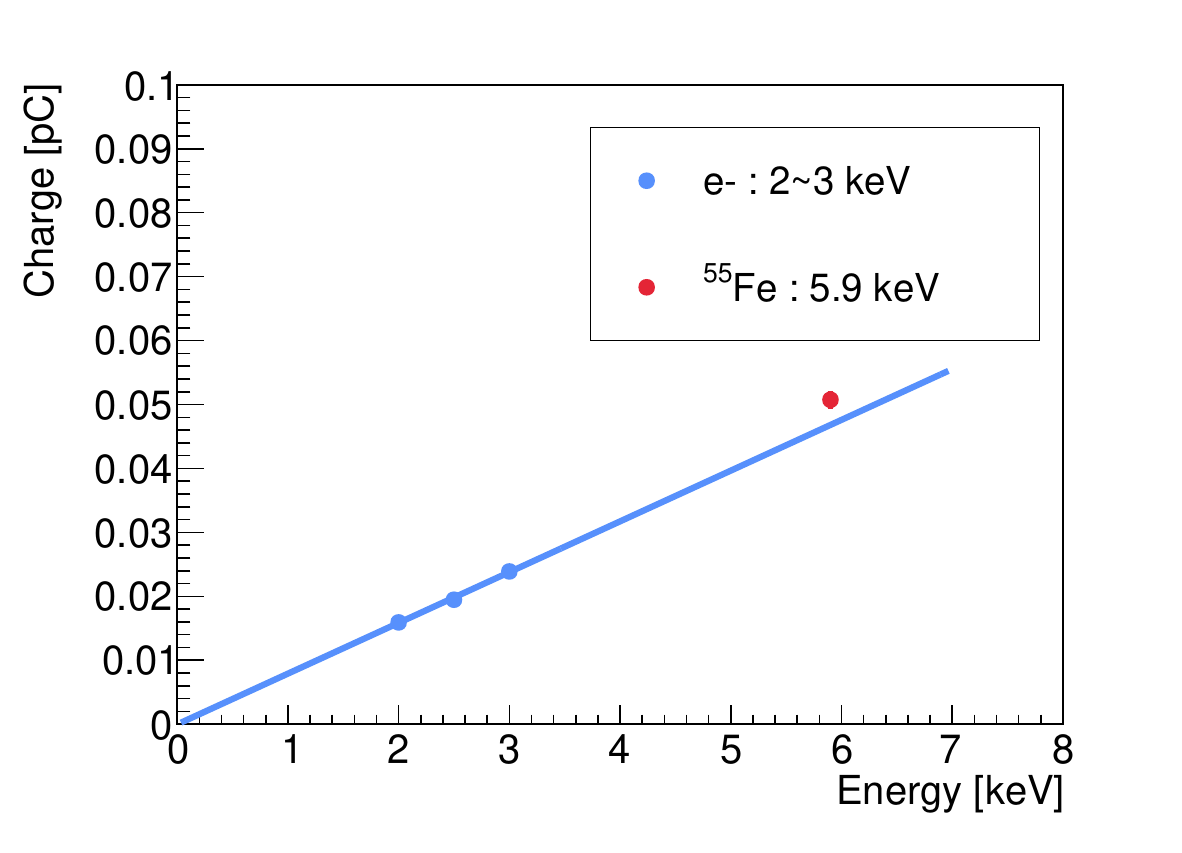}
    \caption{Linearity of the measured charges as a function of the source energies. The measured charges were converted from the pulse heights of the signals. The blue line represents a linear fitting result with electron-gun measurements.}
    \label{fig:linearity}
\end{figure}

The energy scale stability was monitored with 
an $^{55}$Fe X-ray (5.9~keV) source before and after each ion-beam run.
Figure~\ref{fig:55Fe_calib} shows the energy spectrum with the  irradiation of X-rays from the $^{55}$Fe source, where the energy scale was calibrated by the electron-gun measurements.
The peak energy corresponding to 5.9~keV was determined by the fitting with a sum function of exponential and gaussian.
Figure~\ref{fig:55Fe_calib} indicates that there is a small ($<10\% $) discrepancy of measured energies between the electron-gun and $^{55}$Fe X-ray measurements.
Figure~\ref{fig:linearity} shows the measured charges as a function of their source energies.
The measured charges are converted from the signal pulse hight, calibrated by test pulse injection to the amplifier.
The residual of the energy scale between the $^{55}$Fe and the electron-gun measurements was 7.7\%.
The residual was corrected in each energy scan and assigned as a systematic uncertainty in the ionization yield measurement.

In order to compensate a limited dynamic range of the readout system, the gas gain was adjusted by controlling the anode voltage.
The anode voltage was set to be $+$1085~V and $+$1000~V in the scan ranges of 5~$\sim$~15~keV and 20~$\sim$~50~keV, respectively.
The gas gains were 3760 and 1170 with anode voltages of $+$1085~V and $+$1000~V, respectively.

\subsection{Ionization yield measurement}
\label{sec:ion_yield}

\begin{figure}[ht]
    \centering
    \includegraphics[width=0.7\textwidth]{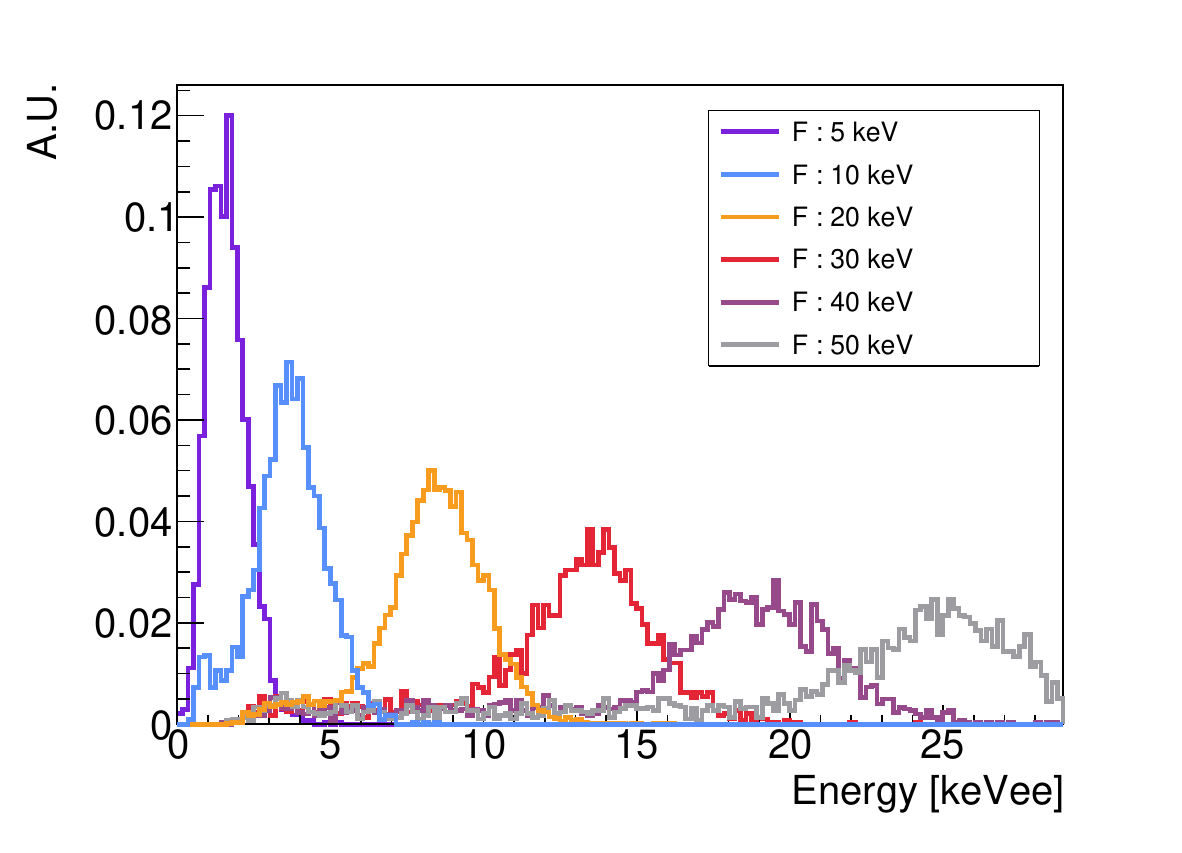}
    \caption{Energy spectra in the F$^{+}$ beam experiment for each scan. The integrals of the spectra are normalized to unity.}
    \label{fig:beam_energy}
\end{figure}

Figure~\ref{fig:beam_energy} shows the energy spectra in the F$^{+}$ beam experiment for each measurement.
The energy scale was calibrated with the electron-gun measurement.
The beam-origin peaks were clearly observed in all measurements.
Although the low energy tales were seen in all spectra due to events passing through the edge of the tapered injection hole, this contribution was found to be negligible.
Therefore the measured energy was determined by fitting the spectrum with a simple gaussian function.
The correlation between the beam energies and measured ones is observed.
With 10~keV and 40~keV F$^{+}$ beams, the measured energy (resolution) was 3.7~keV$_{\rm ee}$ (29\%, sigma) and 18.6~keV$_{\rm ee}$ (12.5\%, sigma), respectively.

\begin{figure}[ht]
    \centering
    \includegraphics[width=0.7\textwidth]{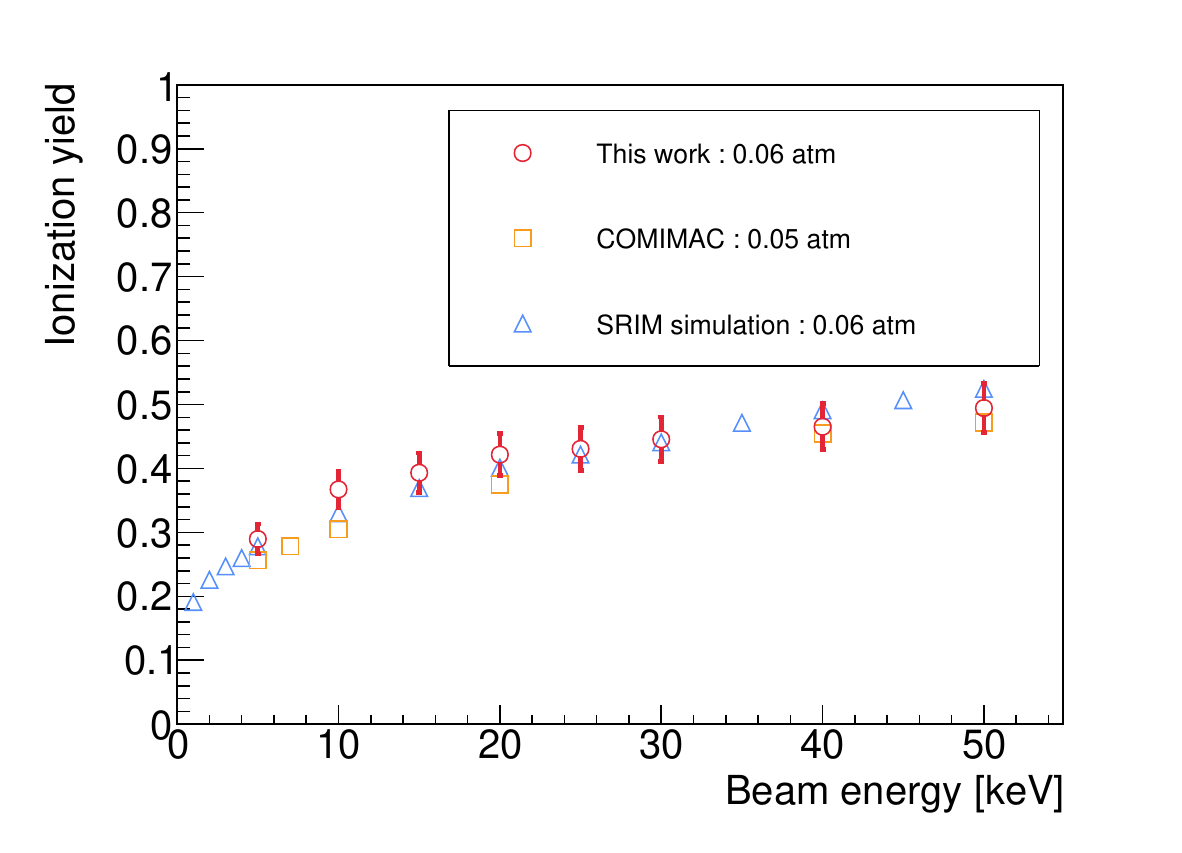}
    \caption{Ionization yield for F$^{+}$ ion at low pressure CF$_{4}$ gas (red circles). Results of the measurement at COMIMAC facility~\cite{Guillaudin:2011hu} and SRIM~\cite{ZIEGLER20101818} simulation are also shown by yellow squares and blue triangles, respectively. Note that the error bar is not shown in the original COMIMAC reference and thus not shown in this figure.}
    \label{fig:ionization_yield}
\end{figure}

The ionization yield is defined as
\begin{equation}
    {\rm ionization \ yield} = \frac{{\rm measured\ energy\ [keV_{ee}]}}{{\rm beam\ energy\ [keV]}}.
\end{equation}
Figure~\ref{fig:ionization_yield} shows the ionization yield as a fuction of the F$^{+}$ beam energy.
Error bars in each plot contain statistical and systematic uncertainties. Here only the energy scale uncertainty is considered for the systematic uncertainty as a main contributor. 
The ionization yield was measured to be 0.45 at 30~keV and slightly grows as the beam energy increases. 
The plot also shows the results of the measurement at the COMIMAC facility~\cite{Guillaudin:2011hu} and SRIM~\cite{ZIEGLER20101818} simulation with the CF$_{4}$ gas pressure at 0.05~atm and 0.06~atm, respectively.

The measured ionization yield was consistent with the prior study from COMIMAC facility and SRIM simulation, although non-negligible systematic uncertainty was assigned in our result.
This uncertainty arises from the discrepancy of measured energies between the electron-gun and the $^{55}$Fe measurements as discussed in Section~\ref{sec:calibration}.
Energies of injected charged particles from both electron-gun and ion beamline were such low  that the energy depositions took places nearby the injection beam hole.
In contrast, energy depositions in the $^{55}$Fe X-ray source measurement occurred anywhere in the cylindrical chamber.
Therefore, the possible reason of this discrepancy is presumed to be a drop of an ionized electron collection efficiency nearby the beam injection hole, although the electrical field in the cylindrical chamber was formed by the copper mesh.
To clarify the source of the discrepancy, measurements with other detectors should be carried on.

We obtained the first data of the low-energy ion injection experiment into a gaseous detector in the low-energy ion beamline at Kanagawa University.
Consequently, the low-energy ion injection mechanism for gaseous detectors was established in this facility. 
Further measurements using various detectors, injection ions and gases are important to understand the behavior of the energy deposition. 

\section{Conclusions}
\label{sec:conclusion}

The measurement of the ionization yield for fluorine ion into the low pressure CF$_{4}$ gas (0.06~atm) was taken place in the low-energy ion beam facility at Kanagawa University.
Low-energy fluorine ion beams with an energy range of 5~$\sim$~50~keV were injected into a dedicated cylindrical wire chamber.
The ionization yield was obtained to be 0.45 at 30~keV, showing a moderate dependence on the ion energy.
This is the first result of the ionization yield measurement for gaseous detector in the beam facility at Kanagawa University.

\appendix

\acknowledgments

This work was partially supported by the Japanese Ministry of Education, Culture, Sports, Science and Technology, Grant-in-Aid (24H02241, 24K00652, 24K07061, 24K21202, 24KK0067, 25H00644, and 25K01025).


\bibliographystyle{JHEP}
\bibliography{main}

\end{document}